\documentclass[12pt]{article}\usepackage{graphicx}
\begin{document}

\title{Introduction to Cosmic F- and D-Strings}


\author{Joseph Polchinski\\[6pt]
\it \normalsize{ Kavli Institute for Theoretical Physics}\\
\it \normalsize{University of California}\\
\it   \normalsize{Santa Barbara, CA 93106-4030}}
\maketitle

\begin{abstract}
In these lectures I discuss the possibility that superstrings of cosmic length might exist and be observable.  I first review the original idea of cosmic strings arising as gauge theory solitons, and discuss in particular their network properties and the observational bounds that rule out cosmic strings as the principal origin of structure in our universe.  I then consider cosmic superstrings, including the `fundamental' F-strings and also D-strings and strings arising from wrapped branes.
I discuss the conditions under which these will exist and be observable, and ways in which different kinds of string might be distinguished.  We will see that each of these issues is model-dependent, but that some of the simplest models of inflation in string theory do lead to cosmic superstrings.  Moreover, these could be the first objects seen in gravitational wave astronomy, and might have distinctive network properties.  

Lectures presented at the 2004 Cargese Summer School.  The outline of these lectures follows hep-th/0410082, but the treatment is more detailed and pedagogical.
\end{abstract}

\newpage
\baselineskip 16pt

\section{Introduction}
Seeing superstrings of cosmic size would be a spectacular way to verify string theory.  Witten considered this possibility in the context of perturbative string theory, and found that it was excluded for several reasons~\cite{Witten:1985fp}.  Perturbative strings have a tension close to the Planck scale, and so would produce inhomogeneities in the cosmic microwave background far larger than observed.  The scale of this tension also exceeds the upper bound on the energy scale of the inflationary vacuum, and so it would have been difficult to produce these strings after inflation, while any strings produced earlier would have been diluted beyond observation.  Ref.~\cite{Witten:1985fp} also identified instabilities that would prevent long strings from surviving on cosmic time scales.

In recent years we have understood that there are much more general possibilities for the geometry of the compact dimensions of string theory, including localized branes, and this allows the string tension to be much lower, anything between the Planck scale and the weak scale.  Also, we have found new kinds of extended object in string theory. Thus the question of cosmic superstrings (and branes) must be revisited, and this has been done beginning in~\cite{Jones:2002cv, Sarangi:2002yt}.  A necessary set of conditions is:
\begin{enumerate}
\item
The strings must be {\it produced} after inflation.
\item
They must be {\it stable} on cosmological timescales.
\item
They must be {\it observable} in some way, but not already excluded.  
\end{enumerate}
Ref.~\cite{Witten:1985fp} thus showed that perturbative strings fail on all three counts.  If we do find models that satisfy these three conditions then there is one more condition that we would hope to satisfy:
\begin{itemize}
\item[4.] Cosmic superstrings should be {\it distinguishable} from other kinds of cosmic string, in particular gauge theory solitons.
\end{itemize}
We will see that each of these four issues is separately model-dependent but that some fortuitous things do happen, and there are simple models in which all these conditions are met, including some of the most fully-developed models of inflation in string theory.  

These lectures give a pedagogical introduction to this subject.  Section~2 reviews the story of cosmic strings that arise as gauge theory solitons.  I discuss the network properties of these strings, and their observational signatures and bounds.  Section~3 discusses the cosmic string candidates in various compactifications of superstring theory.  Sections~4 through~7 go through each of the conditions on the list above, explaining the issues and their dependence on the details of the compactification.  Section~8 presents brief conclusions.

The outline of these lectures (and some of the prose!) follows~\cite{Polchinski:2004hb}, but the treatment is more detailed and pedagogical.

\section{Cosmic string review}

\subsection{String solitons}

Cosmic strings might also arise as solitons in a grand unified gauge theory, and for some time these were a candidate for the source of the inhomogeneities that led to the structure in our universe.  I start with a review of this subject, since almost all of the ideas and results carry over to superstrings.  A complete treatment of this subject would require at least a full volume, and the interested reader is directed to the book by Vilenkin and Shellard~\cite{VilShell} 
and the review by Hindmarsh and Kibble~\cite{Hindmarsh:1994re}; for a recent look at the subject see~\cite{Kibblenew}.  I will have space only to give an overview of the most important ideas.

In any field theory with a broken $U(1)$ symmetry, there will be classical solutions that are extended in one dimension~\cite{Abrikosov:1956sx, Nielsen:1973cs}.\footnote{I will make a more precise statement, in terms of homotopy groups, in section~4.}  Consider the standard symmetry-breaking potential for a complex scalar field,
\begin{equation}
V(\Phi) = \lambda ( |\Phi|^2 - v^2)^2\ . \label{potent}
\end{equation}
This has a ring of degenerate minima, $\Phi = v e^{i \psi}$ for any phase $\psi$.  Suppose we have a configuration in two space dimensions such that the potential energy is nonzero only in a localized region, falling off rapidly at infinity.  At sufficiently large radius $R$ the field must then 
be approximately at a minimum of the potential, but possibly a different minimum in different directions:
\begin{equation}
\Phi(R,\theta) \approx v e^{i\psi(\theta)}\ .
\end{equation}
As we circle the origin at long distance, the phase $\psi(\theta)$ must make some integer number $n$ of circuits of the ring of minima.  This winding number is conserved in time because it is
determined by the topology of the field at arbitrarily long distance.

Thus there is a conserved topological quantum number, and the minimum energy configuration in each sector is a stable topological soliton.  By continuity, whenever the winding number $n$ is nonzero the field must somewhere pass through zero, so a nonzero winding number implies that there must be a nontrivial `lump' of potential energy.  The size of this core region is set by the Higgs mass $m_{H} \sim a\sqrt\lambda$.  Quantum corrections shift the minimum energy in each sector, but do not destabilize the soliton unless they actually restore the $U(1)$ symmetry.  
Assuming rotational symmetry gives the form
\begin{equation}
\Phi(r,\theta) = f(r) e^{i n \theta}\ ,\quad f(0) = 0\ ,\quad f(\infty) \to v\ . \label{wind}
\end{equation}
We can add in the third dimension simply by extending the polar coordinates to cylindrical coordinates $(r,\theta,z)$ with the field independent of $z$, so that the lump becomes a long straight string.

The broken $U(1)$ can be either a global or a gauge symmetry.  In the global case the Hamiltonian density at long distance is
\begin{equation}
{\cal H} = |\Pi|^2 + | \vec\partial \Phi|^2 + V(\Phi)
\to \frac{v^2 n^2}{r^2}\ ,\quad r\to\infty\ ,
\end{equation}
because the gradient of $\Phi$ falls as $1/r$.
The string tension, $\int dr\, 2\pi r {\cal H}$, is logarithmically divergent at long distance.  Nevertheless these solutions are of interest.  For example, for two oppositely oriented straight strings separated by a distance $L$, the divergence is cut off at $L$.  This configuration is topologically trivial and unstable to annihilation of the long strings, but the force between them falls as $1/R$ so in fact the strings can be quite long-lived.

When the $U(1)$ is gauged, the gradient in the covariant derivative $\vec D \Phi = \vec \partial \Phi - i \vec A \Phi$ can be cancelled by an appropriate gauge field,
\begin{equation}
\vec A(r,\theta) = \hat\theta g(r)\ ,\quad g(0) = 0\ ,\quad g(\infty) \to  \frac{n}{r}\ .
\end{equation}
The energy is then finite.  The core of the gauge string contains a magnetic flux
\begin{equation}
\int F = \oint_C A = 2\pi n\ ,
\end{equation}
where $C$ is a large circle around the origin.  Most of the literature on cosmic strings deals with gauge strings.  Most of the superstrings that we will encounter will indeed be analogous to gauge strings, and so this is what we assume in the discussion of network properties and signatures below, but we will see that there is also a possibility of global strings.

\subsection{Formation of strings}

These classical string solutions exist whenever there is a broken $U(1)$ symmetry, and whenever a $U(1)$ symmetry {\it becomes} broken during the evolution of the universe a network of strings must actually form.  Such a symmetry-breaking transition can occur as a result either of coupling to a time-dependent scalar field or of thermal effects.  In the first case the potential for $\Phi$ might be of the form
\begin{equation}
V(\Phi) = \lambda ( |\Phi|^2 - g \chi )^2
\end{equation}
with $\chi$ a scalar field that is slowly rolling from negative values, for which the potential is minimized at $\Phi = 0$, to positive values, where the symmetry is broken.  In the thermal case, the vacuum is found by minimizing the free energy $E - TS$.  The entropy is usually larger in the unbroken phase, because it has a greater number of massless degrees of freedom.  Thus the unbroken phase might be favored at high temperature and then the symmetry break as the universe cools and we approach the zero-temperature potential~(\ref{potent}).

The Higgs field $\Phi$ thus starts at zero, and then when the symmetry breaks it rolls down to one of the vacua.  However, it will not role in the same direction everywhere: by causality, it cannot be correlated on distances greater than the horizon scale.  In practice, the correlation length is usually less than this, determined by the intrinsic length scales of the field theory.
Thus it chooses random directions in different places, and inevitably there will be some trapped winding, so that strings are left over at the end.  This is the Kibble argument~\cite{Kibble:1976sj}, and indeed it is what simulations show.  A fraction of the string is in the form of finite sized loops, and a fraction is in the form of infinite strings; the latter enter and leave the boundaries of the simulation volume no matter how large this is taken to be.  These populations are cleanly separated because the distribution of lengths of finite loops falls rapidly for long loops, giving a convergent integral for the total amount of string in loops.  Presumably the existence of the infinite strings is implied by the causality argument. They are important because they begin to stretch with the expansion of the universe, while the small loops quickly decay away.

\subsection{Evolution of string networks}

To get some understanding of the evolution of the string networks, let us first make the crude assumption that the string network just expands along with the growth of the universe.  If we look at a given comoving region, the length of string within, and therefore the total energy, grows as the scale factor $a(t)$.  The comoving volume scales as $a(t)^3$, so the energy density $\rho_s$ scales as $a(t)^{-2}$.  This would dominate over matter, $\rho_m \propto a(t)^{-3}$, and radiation, $\rho_r \propto a(t)^{-4}$.

However, there are several processes that reduce the string energy density relative to this estimate:\footnote{For reasons of space I will focus on strings whose only long-distance interactions are gravitational.  Global strings would also have a long-ranged Goldstone boson field, and could decay via Goldstone boson emission.  There are also superconducting strings~\cite{Witten:1984eb}, which have strong couplings to gauge fields.  These are perhaps less likely to arise in the superstring case, for reasons that I will explain.}  
\begin{enumerate}
\item
The long strings, which form as random walks, expand uniformly only on scales greater than the horizon length $t$.  On shorter scales they tend to straighten over time.
\item
When two strings collide they can either pass through each other or they can reconnect (intercommute), as in Fig.~1.
\begin{figure}[t]
\begin{center}
\includegraphics[height=.13\textheight]{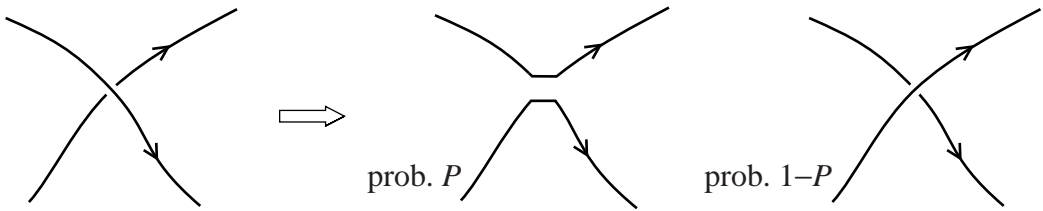}
\caption{When two strings of the same type collide, they either reconnect, with probability $P$, or pass through each other, with probability $1-P$.  For classical solitons the process is deterministic, and $P=1$ for the velocities relevant to the string network.  When the strings reconnect, a sharp kink is left on each new string, and each of these separates into a right-moving and a left-moving kink: the result is four kinks, as shown.}
\end{center}
\end{figure}
For gauge theory solitons, the value of $P$ is essentially 1.  In an adiabatic collision, gauge theory strings always reconnect, $P=1$, because reconnection allows the flux in the string core to take an energetically favorable shortcut.  Simulations show that this remains true at the moderately relativistic velocities that are present in string networks~\cite{Shellard:1987bv, Matzner, Moriarty:1988fx}.\footnote{  Cosmic strings move at moderately relativistic speeds.  In flat spacetime the virial theorem implies that $v^2$ has a mean value of $\frac12$; this is somewhat reduced by the expansion of the universe.}
\begin{enumerate}
\item
When a long string intersects itself, a loop breaks off.  Loops on scales shorter than the horizon do not expand and so behave like massive matter.
\item
When two different long strings reconnect they produce two long kinked strings.  The kinks tend to straighten in time, as in point 1, which reduces the total string length.  (This effect is not enough to produce a scaling solution without also closed loop formation from self-intersection).
\end{enumerate}
\item
The closed loops eventually decay through gravitational radiation.
\end{enumerate}

Simulations show that these processes act with the maximum efficiency allowed by causality, so that the appearance of the string network at any time looks the same when viewed at the horizon scale $t$,  with a few dozen strings spanning the horizon volume and a gas of loops of various sizes~\cite{Albrecht:1989mk, Bennett:1989yp, Allen:1990tv}, as in Fig.~2.  
\begin{figure}[t]
\begin{center}
\includegraphics[height=.40 \textheight]{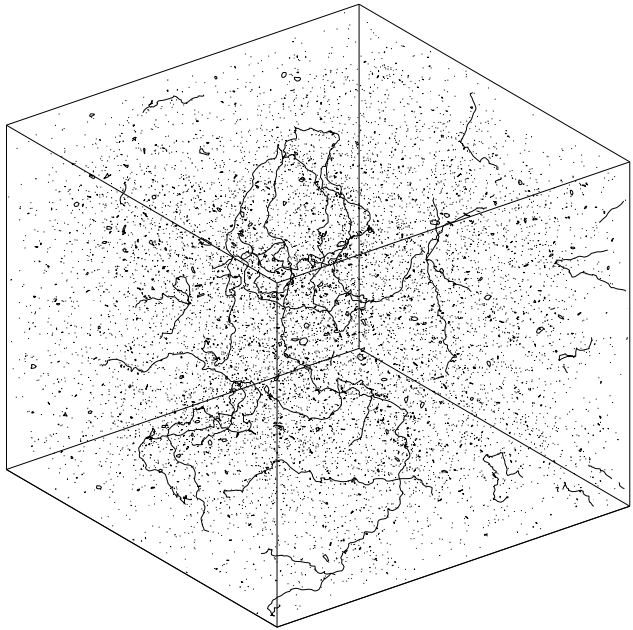}
\caption{Cosmic string simulation.  A side of the cube is around a third of the horizon length.    Strings that appear to end are just leaving the simulation volume.  From Allen and Shellard~\cite{Allen:1990tv}.}
\end{center}
\end{figure}
This is known as the {\it scaling solution}.  The scaling solution is an attractor: if we start with too much string, the higher collision rate will reduce it, while if we start with too little then there will be few collisions until the amount of string per horizon volume (which is increasing relative to the comoving volume) approaches the scaling value.  Thus the details of the initial distribution do not matter, as long as there are some infinite strings.

The total length of string within a horizon is then a numerical constant times $t$, and the horizon volume is $t^3$, so the energy density $\rho_s$ is a constant times $\mu t^{-2}$, where $\mu$ is the string tension.  The scale factor is proportional to $t^{1/2}$ in the radiation-dominated era and $t^{2/3}$ in the matter  dominated era, so it works out that the ratio of the string energy density to the dominant energy density in each era is a constant.  The simulations show that $\rho_s/\rho_m \sim 60 G\mu$ during the matter-dominated era and $\rho_s/\rho_r \sim 400 G\mu$ during the radiation dominated era.  This is our first encounter with the dimensionless combination $G\mu$, about which we will have much more to say, but for now we just assert that we are interested in values of $G\mu$ less than $10^{-6}$.  The strings would thus be a small but fixed fraction of the total density.  In particular, it would take some very exotic network scenario for strings to be the dark matter.

The horizon distance, approximately $t$, sets the overall scale of the string network, but there are important features on smaller length scales.  In particular, there is a lower cutoff on the size of loops, coming from their decay via gravitational radiation.  Dimensionally, the gravitational wave power emitted by a loop of size $l$ is $G \mu^2$.  Equating the loop mass $\mu l$ to the power times $t$ determines the typical lifetime for a loop of size $l$; equivalently it determines the size $l$, as a function of the time $t$, below which the decay becomes rapid.  Inserting the constants from a more careful treatment, the result is
\begin{equation}
l \sim 50 G\mu t\ . \label{loopsize}
\end{equation}
Taking $G\mu = 10^{-9}$, which will turn out to be a typical value, gives $l \sim 500$ lightyears today.  Thus there is large hierarchy of scales in the network.

Another important property of the network is the existence of a kinked structure at short distance~\cite{Bennett:1989yp}.  We have noted that reconnection of loops produces kinks.  As $t$ increases the number of kinks per horizon length necessarily increases, and this effect is stronger than the tendency of the kinks to straighten as the string stretches.  The study of this structure in simulations is limited by the resolution of the simulations, but it has generally been assumed that it is cut off by gravitational radiation at the same scale~(\ref{loopsize}) (see however~\cite{Siemens:2002dj}).  The short distance structure is important because the simulations indicate that it determines the typical size of the loops that break off the infinite strings.  

In summary, we should note that the behavior of string networks is challenging to understand in detail: the large ratio of scales makes numerical simulation difficult, and while there exist various analytic models, the problem is very nonlinear and existing models only capture part of the physics.  If cosmic strings are ever detected, it will become important to work toward a much more precise understanding.

Finally, it is interesting to contrast the properties of one-dimensional defects with those of two-dimensional and zero-dimensional defects.  If a network of domain walls forms and subsequently annihilates to the maximum extent allowed by causality, then the typical spacing between walls will be of order $t$.   The energy density in domain walls will then scale as $1/t$, as compared to the $1/t^2$ of strings, and this would come to dominate over the matter and radiation densities.  Such a network is therefore excluded.  For point particles, annihilation is not able to reduce the density to the scaling value: points have a harder time finding each other than strings.  The spacing then scales as $a(t)^{-1}$ and the density as $a(t)^{-3}$, just like other massive matter; such defects would therefore come to dominate in the radiation-dominated era and are excluded.

Thus, whereas strings represent a cosmic opportunity, domain walls and point defects are forbidden.  In section~4 we will discuss potential instabilities that could eliminate cosmic strings, and the analogous decays would also provide solutions to the domain wall and point defect problems.  However, there is one kind of point defect that provides a serious challenge, namely the magnetic monopole~\cite{Zeldovich:1978wj,Preskill:1979zi}.  These exist in all theories that unify electromagnetism with the other interactions~(for a recent overview see~\cite{Polchinski:2003bq}).  They cannot decay (magnetic charge conservation) and they are not confined (at least today), so they present a vexing cosmological problem.  However, if the phase transition in which they are produced took place before inflation, they would have been diluted the point that a monopole would probably never be seen; this was one of the original motivations for inflation~\cite{Guth:1980zm}.  By the same token, cosmic strings will only be of interest if they are produced at the end of inflation, or later.

\subsection{String signatures and bounds}

Since we are considering strings whose only interactions are gravitational, all of their effects are controlled by the dimensionless product of Newton's constant and the string tension, $G\mu$.  One can think of this in two ways: it is the string tension in Planck units, and it sets the size of the typical metric perturbation produced by a string.  For example, the geometry around a long straight string is conic, with a deficit angle $8\pi G\mu$.  

The string network produces inhomogeneities proportional to $G\mu$, and because of the scaling property of the string network these are scale invariant.  A value $G\mu \sim 10^{-5.5}$ would give scale-invariant perturbations of the right magnitude to produce the galaxies and CMB fluctuations.\footnote{I am going to quote values of $G\mu$ to the nearest half order of magnitude.  This is less precise than most numbers given in the literature, but it is all I will need and it  roughly reflects the uncertainties in the understanding of the string network.  To give more precise numbers would require a more detailed discussion, and so the reader should consult the references.} 
This was a viable theory for some time~\cite{Zeldovich:1980gh, Vilenkin:1981iu}, but it is now excluded.
For example, the CMB power spectrum is wrong: the actual spectrum shows a pattern of peaks and dips, whereas the spectrum from strings would be smooth.  There is a simple reason for this.  The fluctuations produced in inflation have a definite phase.  This phase is maintained from the end of inflation until the perturbations go nonlinear, and is imprinted as oscillations of the power spectrum.  Strings, on the other hand, each keep their own time, there is no common phase.  Fitting cosmological constant plus cold dark matter {\it plus} strings to the CMB power spectrum gives an upper limit $G\mu< 10^{-6}$~\cite{Pogosian:2003mz, Pogosian:2004ny}.  Beyond the power spectrum, strings will produce nongaussianities in the CMB.  Recent limits are somewhat stronger than those from the power spectrum, around $G\mu< 10^{-6.5}$~\cite{Jeong:2004ut,Lo}.\footnote{This limit is for a network of strings.  For a single stray cosmic string, the limit is an order of magnitude weaker.}
Incidentally, strings affect the CMB in more than one way.  Strings at early times will induce the matter inhomogeneities that are reflected in the CMB, but even if the CMB were uniform, strings at the present time would produce apparent inhomogeneities.  We have noted that a static string bends light.  If a string moves with velocity $v$ transverse to the observer there will be a differential redshift of order $8\pi v G \mu$ between the two sides.

Another limit on $G\mu$ comes from pulsar timing.  Because the energy in the strings eventually goes into gravitational waves, strings produce a large stochastic gravitational wave background.  The classic reference on this subject~\cite{Kaspi:1994hp} quotes a limit on the energy density in stochastic gravitational waves, per logarithmic frequency range, as $\Omega_{\rm GW} < 1.2 \times 10^{-7}$, using $h^2 = 0.5$ for the Hubble parameter.  In the frequency range of interest one can estimate the stochastic background from a network of strings as $\Omega_{\rm GW} = 0.04 G\mu$~\cite{VilShell}, meaning that $G\mu < 10^{-5.5}$.  Different analyses of {\it the same} data, with different statistical methods, give a (controversial) bound a factor of 6 stronger~\cite{Thorsett:1996dr} and another a factor of 1.6 weaker~\cite{McHugh:1996hd}.  The limits from pulsars should increase rapidly with greater observation time, and recent work~\cite{Lommen:2002je} quotes
a bound 30 times stronger than~\cite{Kaspi:1994hp}, meaning that $G\mu < 10^{-7}$.  However, this paper again obtains much weaker limits using other methods and so this should not be regarded as a bound until there is agreement on the analysis.\footnote{I would like to thank E. Flanagan and H. Tye for communications on this point.}  
The bounds here are from gravitational waves with wavelengths comparable to the size of the emitting string loop, and do not include a potentially substantial enhancement due to high-frequency cusps; see section~6.

Thus far we have quoted upper bounds, but there are possible detections of strings via gravitational lensing.  A long string will produce a pair of images symmetric about an axis, very different from lensing by a point mass.  Such an event has been reported recently~\cite{Sazhin:2003cp, Sazhin:2004fv}.  The separation of around two arc-seconds corresponds to  $G\mu$ equal to $4 \times 10^{-7}$ times a geometric factor of order 1.  I have heard varying opinions on how seriously to take such an observation, as similar pairs in the past have turned out to coincidental.  There is further discussion in the review~\cite{Kibblenew}, which also discusses a possible time-dependent lens, as from an oscillating loop, with $G\mu \sim 10^{-7.5}$~\cite{Schild:2004uv}.

\section{Cosmic strings from string theory}

Now let us consider the candidates for cosmic superstrings that might arise in various vacua of string theory, starting with the perturbative heterotic string.  The gravitational and gauge couplings in this case have a common origin from the closed string interaction, and so by calculating the trilinear amplitudes one finds the relation (e.g. Eq.~18.2.4 of~\cite{Polchinski:1998rr})
\begin{equation}
4 \kappa^2 = \alpha' g^2\ .
\end{equation}
The gauge and gravitational couplings are equal up to a factor of $\alpha'$ required by the dimensions, and a factor of 4 which arises from various conventions.  This holds for the ten-dimensional couplings, but because the gauge and gravitational fields both live in the bulk they reduce in the same way, $\kappa_4^2 = \kappa_{10}^2/V$ and $g_4^2 = g_{10}^2/V$ where $V$ is the compactification volume.  Expressing the relation in terms of $G = \kappa_4^2/8\pi$, $\mu = 1/2\pi\alpha'$, and the heterotic gauge coupling $\alpha_{\rm h} = g_4^2/4\pi$ gives
\begin{equation}
G\mu = \frac{\alpha_{\rm h}}{16\pi}\ .
\end{equation}
The heterotic gauge coupling should be at or above the minimal GUT value $1/20$, so $G\mu$ is at least of order $10^{-3}$.  The existence of such cosmic strings is therefore excluded~\cite{Witten:1985fp}.

We cannot turn this around and use it as evidence against perturbative heterotic string theory, because in fact a cosmic string network of this type could not form.  The CMB limit on tensor perturbations implies an upper limit on the vacuum energy during inflation, $G^2 V_{\rm inf} \,{\stackrel{<}{{_\sim}}}\, (\delta T/T)^2 \sim 10^{-10}$.  In most cases this bounds the tension of any strings that might subsequently form, $G\mu \,{\stackrel{<}{{_\sim}}}\, (G^2 V_{\rm inf} )^{1/2} \,{\stackrel{<}{{_\sim}}}\, 10^{-5}$.  Thus, perturbative heterotic strings could only have been produced in a phase transition before inflation, and the many $e$-foldings of inflation then make it unlikely that any would be found within our horizon.  There is a second reason that heterotic strings would not reach cosmic size, as we see in section~5.

For the perturbative type I string the dimensional analysis is a bit different but the conclusions are the same.  The {\it strongly} coupled $E_8 \times E_8$ heterotic string can have a lower tension~\cite{Witten:1996mz}.  This is because the gauge fields live on the nine-dimensional boundary while gravity lives in the ten-dimensional bulk.  More generally, if the gauge fields 
are confined to a brane while gravity propagates in the bulk, the string tension is suppressed by some power of $R/L_{\rm P}$ where $R$ is the size of the dimensions transverse to the brane~\cite{Arkani-Hamed:1998rs,Antoniadis:1998ig}.

Even without large compact dimensions, the string tension can be reduced by a gravitational redshift (warp factor)~\cite{Randall:1999ee} 
\begin{equation}
ds^2 = e^{2 A(y)} \eta_{\mu\nu} dx^\mu\, dx^\nu + \ldots
\end{equation}
where $y$ are the compact coordinates.
The string is localized at some point $y_0$ in the compact space, so the string worldsheet action is proportional to
\begin{equation}
\mu = \frac{ e^{2 A(y_0)} }{2\pi\alpha'}\ .
\end{equation}
In interesting compactifications there are often {\it throats} where the warp factor $e^{2 A}$ is much less than 1.  The string will fall to the bottom of such a throat, so the four-dimensional tension $\mu$ measured in four dimensions can be reduced by a large factor relative to the tension $1/2\pi\alpha'$ seen by a local ten-dimensional observer.
This is the RS idea, that different four-dimensional scales arise from a single underlying scale through gravitational redshifting.

Thus the tension is essentially a free parameter, which can lie anywhere from the Planck scale down to the experimental limit near the weak scale.  For example, in the warped models it is determined by the depth of the throat, which comes out as the exponential of a function of flux integers that can take a wide range of values~\cite{Giddings:2001yu}.  

The range becomes much smaller when we focus on {\it brane-antibrane inflation} models~\cite{Dvali:1998pa,Alexander:2001ks,Burgess:2001fx,Dvali:2001fw}.  This is a nice geometric idea for obtaining a slow-roll inflationary potential, and the basis for most current attempts to describe inflation in string theory.  That is, the early universe could have contained an extra brane-antibrane pair, separated in the compact directions.  The potential energy of these branes would drive inflation.  The inflaton field is then the separation between the branes: this has a potential which is rather flat when the branes are separated and steepens as they approach, until at some point a field becomes tachyonic and the brane and antibrane annihilate rapidly.  

To understand inflationary cosmology in detail one needs to know the scalar potential.  This potential has long been problematic, especially for states of positive vacuum energy,
because of potential instabilities in the moduli directions.  Recently the tools have been developed to identify a large class of stable solutions~\cite{Kachru:2003aw, Silverstein:2004id}.  Thus we will focus on the resulting  \mbox{\it K\hspace{-7pt}KLM\hspace{-9pt}MT model}~\cite{Kachru:2003sx}, since it is the most detailed model of inflation in string theory.  This is based on a strongly warped compactification of the IIB string theory, with inflation arising from a D3/${ \hspace{1pt}\overline{\hspace{-.5pt}{\rm D3}\hspace{-1.5pt}}\hspace{1.5pt} }$
pair at the bottom of a throat.

In models of brane inflation, the value of $G\mu$ can be deduced from the observed magnitude of the CMB fluctuations $\delta T/T$.  That is, one assumes that $\delta T/T$ arises from the quantum fluctuations of the inflaton; this is natural given the flat form of the inflaton potential.  For any given brane geometry the inflaton potential has a definite functional form.  For example, in the {\mbox{K\hspace{-7pt}KLM\hspace{-9pt}MT}} model the D3/${ \hspace{1pt}\overline{\hspace{-.5pt}{\rm D3}
 \hspace{-1.5pt}}\hspace{1.5pt} }$ potential is $V \sim V_0[1 - O(\phi^{-4})]$.  One then finds (Eq.~C.12 of~\cite{Kachru:2003sx})
\begin{equation}
G^2 V_0 \sim 0.05 (\delta T/T)^3 N_e^{-5/2} \sim 1.5 \times 10^{-20}\ ; \label{infdep}
\end{equation}
here $N_{e} \sim 60$ is the number of $e$-foldings at the wavelengths responsible for structure formation.
The vacuum energy from a D3/${ \hspace{1pt}\overline{\hspace{-.5pt}{\rm D3}\hspace{-1.5pt}}\hspace{1.5pt} }$ in the warped space is $2 T_3 e^{4A_0}$ where $A_0$ refers to the bottom of the inflationary throat.  The F-strings produced after inflation (the mechanism for which is the subject of the next section) will sit in the same throat and so $\mu = e^{2A_0}/2\pi\alpha'$.  Since $T_3 = T_{\rm F}^2/2\pi g_s$ in ten dimensions~\cite{Polchinski:1998rr} we have for the tension of the IIB string in the inflationary throat
\begin{equation}
\mu_{\rm F} \sim 2 \times10^{-10}  \sqrt{g_{\rm s}}\ ,
\end{equation}
where $g_{\rm s}$ is the string coupling.

One might take $g_{\rm s}$ to have a representative value of $0.1$.  There is one very specific class of models in which it is related to observed quantities.  If the matter fields live on D3's or
 ${ \hspace{1pt}\overline{\hspace{-.5pt}{\rm D3}\hspace{-1.5pt}}\hspace{1.5pt} }$'s, then $g_{\rm s}$ turns out to be exactly $\alpha_{\rm YM}$.  However, $g_{\rm s}$ is really $e^{\Phi}$ at the position of the brane, and the dilaton $\Phi$ in most solutions varies in the compact space and so will not have the same value at the position of the strings as it has at the Standard Model brane.  In the special case that it is constant, we can identify $g_{\rm s}$ in the brane tension with the unified coupling $\alpha_{\rm GUT}$.  Unification in brane models is generally nonstandard, but a likely value for $\alpha_{\rm GUT}$ is still around $0.1$.
 
 Note that these warped manifolds can have many throats and therefore strings of many different tensions.  One possibility is that the Standard Model fields live in a throat whose depth corresponds to the weak scale, whereas the depth~(\ref{infdep}) of the inflationary throat gives something closer to the GUT scale.  Thus there would be the possibility to see TeV-scale strings and extra dimensions at accelerators, and at the same time GUT-scale strings in the sky.

In addition to the fundamental F-strings, string theory has many other extended objects.  For example the IIB theory has a D-string whose tension in ten dimensions is $T_{\rm D} = T_{\rm F}/g_{\rm s}$, and so in the {\mbox{K\hspace{-7pt}KLM\hspace{-9pt}MT}} model
\begin{equation}
\mu_{\rm D} \sim 2 \times10^{-10}/ \sqrt{g_{\rm s}} ;
\end{equation}
the geometric mean $(\mu_{\rm F} \mu_{\rm D})^{1/2}$ is independent even of $g_{\rm s}$.  There are also bound states of $p$ F-strings and $q$ D-strings with a distinctive tension formula
\begin{equation}
\mu_{(p,q)} = \mu_{\rm F} \sqrt{p^2 + q^2/g_{\rm s}^2}\ . \label{eq:tensions}
\end{equation}

There are also a variety of higher dimensional D-, NS- and M-branes in the various regimes of string theory.  Any of these will look like a string if all but one of its spatial directions is wrapped on a compact cycle.  In the {\mbox{K\hspace{-7pt}KLM\hspace{-9pt}MT}}  model, any brane not contained fully in a throat will have a Planck-scale tension because a portion of it passes through the region where the warp factor is close to 1, so we will get additional interesting strings only if there are nontrivial cycles in the throat.  The most generic throat, the Klebanov-Strassler throat~\cite{Klebanov:2000hb} based on the conifold singularity~\cite{Candelas:1989js}, has no cycles that give additional strings, but there are many less generic singularities with a wide variety of nontrivial cycles; these have been extensively studied in the context of gauge/string duality. 

For models based on large compact dimensions, no fully stabilized examples are known.  One can fix the bulk moduli by hand and study inflation based on the brane moduli.  This is done in~\cite{Sarangi:2002yt,Jones:2003da}, which consider a variety of wrapped branes in type II compactifications, and find values in the range
 \begin{equation}
 10^{-12} \,{\stackrel{<}{{_\sim}}}\,G\mu \,{\stackrel{<}{{_\sim}}}\, 10^{-6}\ .
 \end{equation}
We should emphasize again that it is the assumption that the inhomogeneities arise from the quantum fluctuation of the inflaton, rather than some other mechanism, that allow us to tie the tension to observed quantities in brane inflation.  If another mechanism is responsible, the inflation scale and the tension could be much lower.
     
It may still turn out to be the case that our vacuum is well-described by weakly coupled heterotic string theory.  If so, inflation and cosmic strings might simply arise in the effective low energy field theory (this would be true for other compactifications as well).  The possible strings include both the magnetic flux tubes discussed in section~2, and also electric flux tubes that exist in strongly coupled confining theories~\cite{Witten:1985fp}.  The heterotic string, whether weakly or strongly coupled, has the advantage that it more readily makes contact with standard grand unification.  A recent paper~\cite{Jeannerot:2003qv} argues that cosmic strings should be present in a wide class of grand unified inflationary models.

\section{Production of cosmic F- and D-strings}

If these branes are D-branes, then there is a $U(1)$ gauge symmetry on each of the brane and antibrane, and this $U(1)\times U(1)$ disappears when the branes annihilate.  One linear combination of the $U(1)$'s is Higged.  The Kibble argument then applies, so that a network of strings must be left over when the branes annihilate~\cite{Jones:2002cv, Sarangi:2002yt}.  These are D-strings, as one can see by studying the conserved charges~\cite{Sen:1998tt, Witten:1998cd}.  More precisely, if the branes that annihilate are $D(3+k)$-branes, extended in the three large dimensions and wrapped on $k$ small dimensions, then the result is $D(1+k)$-branes that extend in one large dimension and are wrapped on the same small dimensions.  The simplest case is $k=0$, where D3/${ \hspace{1pt}\overline{\hspace{-.5pt}{\rm D3}\hspace{-1.5pt}}\hspace{1.5pt} }$ annihilation produces D1-branes.
          
The second linear combination of $U(1)$'s is confined.  We can think of confinement as dual Higgsing, by a magnetically charged field, and so we would expect that again the Kibble argument implies production of strings~\cite{Dvali:2003zj, Copeland:2003bj}.  These are simply the F-strings, the `fundamental' superstrings whose quantization defines the theory, at least perturbatively.

It is notable that this process produces only strings, and not the cosmologically dangerous monopoles or domain walls~\cite{Jones:2002cv, Sarangi:2002yt}.  The breaking of a $U(1)$ produces defects of codimension two, and the Kibble argument requires that the codimension be in the large directions: the small directions are in causal contact.\footnote{Ref.~\cite{Dvali:2003zj} identifies processes that might suppress production of D-strings.  Ref.~\cite{Barnaby:2004dz} argues that in some cases there can be disorder in the compact dimensions and so monopoles and domain walls would be produced.  We believe that, at least for the {\mbox{K\hspace{-7pt}KLM\hspace{-9pt}MT}} model, neither of these should be relevant.}

It is striking that what appears to be the most natural implementation of inflation in string theory produces strings and not dangerous defects, but we should now ask how generic this is.  Even these models are not, in their current form, completely natural: like all models of inflation they require tuning at the per cent level to give a sufficiently flat potential~\cite{Kachru:2003sx, Iizuka:2004ct}.  There might well be other flat regions in the large potential energy landscape of string theory.  An optimistic sign is that there are arguments entirely independent of string theory to indicate that inflation terminates with a symmetry-breaking transition: this is known as hybrid inflation~\cite{Linde:1993cn}, and leads to efficient reheating as well as production of strings~\cite{Yokoyama:1989pa,Kofman:1995fi,Tkachev:1998dc}.  However, not every symmetry-breaking pattern produces strings.  For example, in strongly coupled heterotic string theory~\cite{Witten:1996mz}, there are M5-branes and M2-branes.  It would seem natural to use M5-branes with two wrapped directions to implement brane inflation.  Since the M2-branes have codimension three relative to the M5-branes it will not be so easy to produce strings; this is currently under investigation.\footnote{ There has also been recent discussion of a more exotic symmetry-breaking pattern in D-brane inflation~\cite{Urrestilla:2004eh,Watari:2004xh,Dasgupta:2004dw}.  For other discussions of string production in brane inflation see~\cite{Halyo:2004zj,Matsuda:2004bk,Matsuda:2004nx}}

In~\cite{Englert:1988th} it was proposed that cosmic fundamental strings could have been produced in a Hagedorn transition in the early universe.  The Hagedorn transition corresponds to the formation of strings of infinite length, and so some percolating strings would survive as the universe cooled below the Hagedorn temperature.  This idea has a simple realization in the warped models.\footnote{Similar ideas are being considered by A. Frey and R. Myers.}  We have noted that the effective string tension, and so the Hagedorn temperature, is different in different throats.  It is possible that after inflation a deeper throat reheats above its Hagedorn temperature, leading to string formation as the universe cools (a black hole horizon would then form at the bottom of the throat, corresponding to the Hawking-Page transition~\cite{Hawking:1982dh,Witten:1998zw}).
In fact, this is essentially equivalent to the Kibble mechanism.  The throat degrees of freedom have a dual gauge description, in terms of which the Hagedorn transition corresponds to deconfinement.  The transition to a confining phase as the universe cools is the electric-magnetic dual of a symmetry-breaking transition.  The strings produced in this way would necessarily have a lower tension than those produced directly in brane annihilation, because of the inefficiency of the extra thermal step.

\section{Stability of strings}

\subsection{Field theory strings}

Ref.~\cite{Witten:1985fp} identified two instabilities that would prevent superstrings from growing to cosmic size.  Actually, these same two instabilities exist for field theory soliton strings~\cite{Preskill:1992ck} --- one for global strings and the other for gauge strings --- so let us first discuss them in this context.

In the case of global strings, we have noted that the long-ranged Goldstone boson has gradient energy.  It does not have potential energy at long distance as long as the broken $U(1)$ symmetry is exact: the broken vacua are then exactly degenerate.  However, there are general arguments in string theory that there are no exact global symmetries~\cite{Banks:1988yz,Polchinski:1998rr}.  More generally, the no-hair theorems imply that black holes can destroy global charges, so in any theory of gravity these can not be exactly conserved.
Thus the degeneracy of the vacua will not be exact; the trough in the potential is tilted and so  there is a potential energy cost even at long distance for a field configuration that circles the trough.  This cost is minimized by have the scalar field make its excursion in a kink of finite width, rather than uniformly in angle as in the exact case~(\ref{wind}).  Thus there is a domain wall, with energy proportional to its area, bounded by the string.  

The wall exerts a transverse force on the string and forces it to collapse, as in Fig.~3.
\begin{figure}[t]
\begin{center}
\includegraphics[height=.13\textheight]{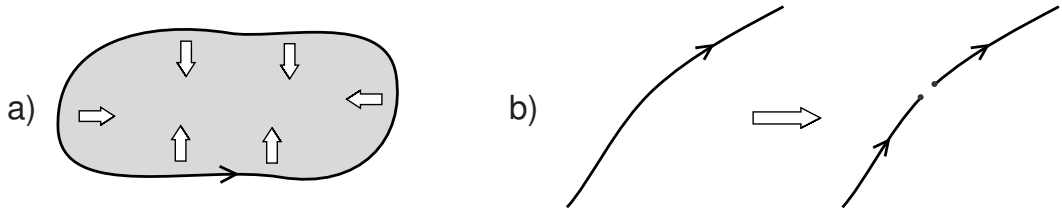}
\end{center}
\caption{Instabilities of macroscopic strings: a) Confinement by a domain wall.  b) Breakage.}
\end{figure}
  This is clear for a loop of string bounding a domain wall, but is less intuitive for a network with infinite strings.  One can picture the network, with the domain walls, as a complicated shape formed from strips whose width is the typical transverse separation between strings.  The timescale for the collapse of the strips, and the disappearance of the strings, is set by the width of the strips, not their (potentially unbounded) length~\cite{Vilenkin:1982ks}.

For gauge strings, the $U(1)$ symmetry is exact because it is a gauge symmetry, and all energies fall exponentially with distance from the string.  A magnetic flux runs along the core of the string, and it is the conservation of this flux that prevents the string from breaking.  However, in any unified theory one expects that there will be electric and magnetic sources for every flux~\cite{Polchinski:2003bq}, so that the string can break by creation of a monopole/antimonopole pair.  If this is possible, it will happen not only once but everywhere along the length of the string, and so the string breaks up into short segments rather thans growing to cosmic length.

There can be absolutely stable strings.  Consider the Abrikosov flux tubes in an ordinary superconductor.  The Higgs field there is an electron pair and has charge $2e$, so the tube has total flux $2\pi/2e$.  However, because there do exist singly charged electrons, Dirac quantization requires the minimum monopole charge to be twice this, $2\pi/e$.  Thus a flux tube cannot end on a monopole, though two can; equivalently one can think of the monopole as a bead on a string, at which the flux reverses.  So the Abrikosov flux tubes (in an infinite system) are absolutely stable.  One can think of this in terms of an unbroken discrete gauge symmetry $(-1)^{Q/e}$, which acts as $-1$ on the electron and $+1$ on the BCS condensate.  As one circles the string, fields come back to themselves only up to this transformation.  Thus the string can be detected by an Aharonov-Bohm experiment at arbitrarily long distance, and so it can never just disappear.  We will refer to these absolutely stable strings as `discrete' strings.  Gauge strings without a discrete gauge charge are truly invisible at long distance, and so there is no obstacle to their breaking.

As an aside, we explain this in terms of the homotopy classification of defects (see e.g. Ref.~\cite{Coleman} for an introduction).  For gauge group $G$ broken to $H$ there is an exact sequence
\begin{eqnarray}
\Pi_2(G/H) &\to& \Pi_1(H) \to \Pi_1(G) \to \Pi_1(G/H)
\nonumber\\
&\to& \Pi_0(H) \to \Pi_0(G) \to \Pi_0(G/H)\ .
\end{eqnarray}
In the discussion of strings we started with $G = U(1)$, for which $\Pi_1(G) = Z$, and broke to $H = I$ for which $\Pi_1(H)=\Pi_0(H)=I$.  Thus the strings are charged under $\Pi_1(G/H) = Z$.  Now however let us embed the $U(1)$ into a semisimple group; we are always free to assume this to be simply connected by going to the covering group, so $\Pi_1(G) = I$ and the stable strings are classified by $\Pi_1(G/H) = \Pi_0(H)_G$ --- that is, by unbroken discrete symmetries that lie in the connected component of the identity in $G$.

It can also be that strings that are not absolutely stable have decays that are slow on cosmic timescales.  For monopole pair creation, for example, the classic Schwinger calculation gives a pair production rate suppressed by $e^{-\pi M^2/\mu}$, where $M$ is the monopole mass.  If the monopole mass is an order of magnitude larger than the scale of the string tension, then the decay will be slow on cosmological time scales.  One can think of a succession of symmetry breakings such as $SU(2) \to U(1) \to 1$.  There are no stable strings for this pattern, but if the scale of the first symmetry breaking is higher than that of the second then there will be string solutions in the effective field theory describing the second breaking, and these can be long-lived.  Similarly the tension of the domain walls might be suppressed by a hierarchy of scales.

\subsection{String theory strings}

\subsubsection{Perturbative strings}

Now let us turn to string theory.  Consider first perturbative strings~\cite{Witten:1985fp}, compactified for example on a torus or a Calabi-Yau manifold.  The type I string couples to no form field, and of course it can break.  The heterotic and type II strings are effectively global strings, because they couple to the long-distance form field $B_{\mu\nu}$ which in four dimensions is dual to a scalar.  A confining force is produced by instantons, which are magnetic sources for $B_{\mu\nu}$ (thus the completeness of the magnetic sources~\cite{Polchinski:2003bq} again enters).  It is simplest to analyze this in terms of the dual scalar $\Theta$, defined by $d\Theta = *dB$ modulo Chern-Simons terms; $\Theta$ is the phase of the earlier $\Phi$.  The instantons become electric sources, meaning that they give rise to explicitly $\Theta $-dependent terms in the effective action: the instanton and anti-instanton amplitudes are weighted by $e^{\pm i \Theta}$. 

It is an interesting exercise to work out the decay rates for the various strings.  The time scale for the breaking of Type I strings is the string scale unless one tunes the string coupling essentially to zero.
For the heterotic string, the one definite contribution comes from QCD instantons.  The field strength for $B_{\mu\nu}$ is of the form
\begin{equation}
H = dB + \omega(A)\ ,
\end{equation}
where $d\omega(A) = {\rm Tr}\,F^2$ and the trace runs over all gauge fields.  Thus the magnetic source $dH$ contains the QCD instanton number.  The scale of the potential is $m_u m_d m_s \Lambda_{\rm QCD} \sim 10^{-7} {\rm GeV}^4 \equiv V_{\rm inst}$.  The potential energy
\begin{equation}
 \int M_{\rm P}^2 (\partial \Theta)^2 + V_{\rm inst} \cos\Theta
\end{equation}
is of order $M_{\rm P}^2 \tau^{-1} + V_{\rm inst} \tau$ for a domain wall of thickness $\tau$. 
This is minimized for $\tau \sim M_{\rm P}/\sqrt{V_{\rm inst}}$ (the inverse axion mass), giving tension
\begin{equation}
T_{\rm dw} \sim M_{\rm P} \sqrt{V_{\rm inst}}\ . \label{dwten}
\end{equation}
The acceleration of a string due to the wall is then $a = T_{\rm dw} /\mu$ which is roughly $\sqrt{V_{\rm inst}} /M_{\rm P}$.  The string network will collapse when this is comparable to the Hubble expansion rate, which is just when the energy density in matter is of order $\sqrt{V_{\rm inst}}$, near the QCD scale.  This is too early to leave observable traces.  Further, the estimate~(\ref{dwten}) is just a lower bound, because it is likely that there are other gauge groups becoming strong at higher energy.  Indeed, it is a challenge to find an scalar (axion) in string theory whose potential is dominated by QCD instantons and so would solve the strong-$CP$ problem.

NS5-brane instantons, wrapped on the whole compact space, are also a magnetic source for $B_{\mu\nu}$~\cite{Callan:1991dj}.  The NS5-brane instantons have action
\begin{equation}
T_5 V_6 = \frac{4\pi^2 \alpha'}{2\kappa_{10}^2} V_6 = \frac{2\pi}{\alpha_{\rm h}} \ .
\end{equation}
This is the same as the action for a gauge theory instanton at the string scale.  In fact the heterotic NS5-branes are just the string-scale limit of instantons, whereas above we considered QCD-scale instantons.
If we take the standard GUT value ${\alpha_{\rm h}} \sim 0.05$ then the contribution to $V_{\rm inst}$ is of order $10^{-50} M_{\rm P}^4 $.  This is larger than the QCD contribution; it might be suppressed by fermion masses (from zero modes), or enhanced if the GUT coupling is increased.

The same estimates apply to perturbative (non-brane) compactifications of the Type II theory, with the gauge fields arising from world-sheet current algebras, although it is known that the  Standard Model cannot be obtained from these~\cite{Dixon:1987yp}.

\subsubsection{Brane models}

In the {\mbox{K\hspace{-7pt}KLM\hspace{-9pt}MT}} model~\cite{Kachru:2003sx} we have noted that the candidate cosmic strings are the F- and D1-strings.  In ten dimensions these strings couple to form fields $B_{\mu\nu}$ and $C_{\mu\nu}$, but these form fields have no four-dimensional massless modes in this model.  They transform nontrivially under the orientifold projection or F theory monodromy that the model requires, and this removes their zero modes.  Thus the strings should be unstable to breakage, and indeed this can occur in several ways.

First, the projection that removes the form field produces an oppositely oriented image string on the covering space of the compactification, and breakage occurs through a segment of the string annihilating with its image.  If the image is coincident with the string, breakage will be rapid.  If the image is not coincident, then the string must fluctuate to find its image.  Due to the warp factor the strings feel a potential
\begin{equation}
\mu = e^{2A(y)} \mu_{10} \ .  \label{eq:tenred}
\end{equation}
The strings are normally confined to the throat where inflation occurs, which as discussed in section~3 has a warp factor $e^{A_0}$ of order $10^{-4}$.  When the string annihilates with its image, the role of the monopole is played by a bit of string that stretches out of the inflationary throat over to the image throat.   Since this passes through bulk region where the warp factor is of order one, the breakage rate is proportional to $e^{-\pi e^{-2A_0}}$ and so totally negligible even on cosmological timescales~\cite{Copeland:2003bj}.  Thus the strings will be stable to self-annihilation if there is no orientifold fixed point (or F-theory equivalent) in the inflationary throat.  There is no particular reason for the throat to be coincident with any fixed point.  Their relative positions are fixed by the complex structure moduli, which depend on flux integers, 
and these are expected to take rather generic values~\cite{Bousso:2000xa,Douglas:2003um,Ashok:2003gk}.

The strings can also break on a brane.  The model must include branes on which the Standard Model (SM) fields live.  If these are D3-branes in the inflationary throat then the strings will break; if they are D7-branes that pass through the inflationary throat then all but the D1-string will break~\cite{Copeland:2003bj}.  If they are outside the throat then the strings are stable for the same reason as above.  In the simplest implementation of the RS idea the SM branes must be in a different throat: the depth of the inflationary throat is something of order the GUT scale, while the depth of the SM throat should be of order the weak scale.  One must ask whether other branes might still be in the inflationary throat, but this is not possible: these would have low energy degrees of freedom which would receive most of the energy during reheating, rather than the SM fields.  So this gives a simple scenario in which the strings are stable, but much more work is needed to see whether it is viable, and whether it is generic.  An important question is whether reheating through the bulk can transfer energy efficiently to the SM throat; a recent paper gives an affirmative answer~\cite{Barnaby:2004gg}.

More complicated geometries with additional cycles in the inflationary throat have additional strings from wrapped branes.  These strings have not been studied in full detail, but in examples one finds that they can be gauge, global, or discrete, depending on the details of the topology (both in the throat and globally).  Similarly their production and stability depends on details.

The stabilization of moduli is not been as fully developed in models based on large compact dimensions, but as a prototype we can consider the $T^6/Z_2$ orientifold which is $T$-dual to the Type I string.  In the Type I string there are BPS D1- and D5-branes~\cite{Polchinski:1995df}, which couple electrically and magnetically to the sole Type I RR form $C_{\mu\nu}$.  There are also non-BPS D3- and D7-branes~\cite{Witten:1998cd}.  All of these give rise to strings when all but one direction are wrapped on the $T^6$.  After $T$-duality these become BPS D7- and D3-branes, and non-BPS D5- and D1-branes, all of which are strings in four dimensions.  The BPS strings couple to RR form fields and so are global.  The relevant instantons come from a Euclidean D-branes.  The calculation is similar to that for the NS5-brane instanton above, with the brane volume replacing $V_6$, but now the compact dimensions can be much larger and so we get essentially stable strings with a long-range RR axion field.

The non-BPS strings separate into two images as in the earlier discussion.  Now, it is important to note that the $T^6/Z_2$ example is nongeneric in an important sense: because of its high degree of symmetry the strings have moduli and move freely in the compact dimensions.  Even if there is not a deep warp factor, supersymmetry breaking will lead to a potential at some scale that will localize the strings.  Just like a large warping, a large separation leads to a large action for the decay and so these strings can again be stable if separated from their images.

Adding the SM and other branes to the large dimension models, the stability of the strings depends on the exact geometry.  In some cases it might appear that a string has both instabilities, in that it couples to a form field but can break on a brane~\cite{Copeland:2003bj,Leblond:2004uc}.  The point is that breakage on a brane implies the existence of a gauge field on the brane to ensure continuity of the form source, and this Higgses the form field so that the result is a gauge string, subject to breakage but not confinement~\cite{Copeland:2003bj}.  We expect that in string theory as in field theory these strings willl be exactly stable only if stabilized by a discrete gauge symmetry visible in the Aharanov-Bohm effect: anything that can happen will~\cite{Polchinski:2003bq}.

In summary, it is encouraging to see that strings can be stabilized as a side effect of certain generic properties such as warping and/or large dimensions, which are needed to lower the inflationary scale below the Planck scale in these models.

\section{Seeing Cosmic Strings}

We will consider below strings that have only gravitational interactions.  However, the possibility, noted above, of global strings with a long-ranged scalar field should not be overlooked.  Another possibility is a strong coupling to SM or other light fields, and in particular superconducting strings~\cite{Witten:1984eb} which carry massless degrees of freedom charged under an unbroken gauge symmetry.
Generically the strings that one finds are rather boring.  In ten dimensions they have massless collective coordinates in the transverse directions, and massless fermions, but these are all gapped by symmetry-breaking effects leaving only the minimal collective coordinates.  Light four-dimensional fields generally live on branes, and we have seen that stability requires cosmic strings to be separated from most other branes, implying interactions of gravitational strength.  An example of an exception would be a D7-brane coincident with a D1 string; the SM fields might live on or couple strongly to the D7-brane.  However, a superconducting string requires that the D1 and D7 be exactly coincident; to ensure this would require a discrete symmetry (which might be orbifolded).  More interesting strings are thus possible in special cases.

The CMB and pulsar bounds on $G\mu$ quoted in section~2 are at the upper end of the brane inflation range, ruling out the highest-tension models.  Both bounds will improve in the coming decade by at least one or two orders of magnitude.  An additional exciting prospect comes from LIGO~\cite{Damour:2000wa,Damour:2001bk}.  Under most circumstances LIGO is at a disadvantage looking for cosmological backgrounds because these fall with increasing frequency: LIGO is looking at frequencies that are $10^{10}$ times those of the pulsar measurements (100 Hz versus years$^{-1}$).  For example, the stochastic background coming from the low harmonics of cosmic strings has a constant energy density per logarithmic scale.\footnote{In making this statement we must consider the relative redshifting of modes at different frequencies.  At both the pulsar and the LIGO frequencies the current bounds come from waves emitted during the radiation dominated era, and in this case the effective red shifts are the same due to the scaling property.  (One can deduce the time of emission for a current freqency $\omega$ by solving 
$\omega = l^{-1}(t) / (1+z(t))$ with $l(t)$ the loop size~(\ref{loopsize}) and $z(t)$ the redshift.)
Lower frequency waves emitted during the matter-dominated era will experience less redshifting.}
This translates into $\omega^2 \tilde h(\omega)^2 \propto \omega^{-1}$ for the Fourier transform $\tilde h$ of the strain (fractional change in the metric).  For a stochastic background (different frequencies uncorrelated) the strain from approximate frequency $\omega$ is of order $\omega^{1/2} \tilde h(\omega)$ which is of order $\omega^{-1}$ here.  Pulsar timing is sensitive to strains around $10^{-14}$ or $10^{-15}$ while LIGO is sensitive to much smaller strains $10^{-22}$ or $10^{-23}$, but the frequency penalty of $10^{10}$ more than offsets this.

However, something unexpectedly nice happens.  When a loop of string in three space dimensions oscillates as governed by the Nambu action, it typically forms a cusp several times per oscillation~\cite{Turok:1984cn}.  To see this go to conformal gauge, where the solution is just a sum or right- and left-movers
\begin{equation}
X^\mu(\sigma,\tau) = f^\mu(\sigma - \tau) + g^\mu(\sigma + \tau)\ ,
\end{equation}
and the gauge conditions imply that $f'$ and $ g'$ are both null.  Classically there is enough residual gauge freedom to set $X^0 = 2\tau$ so that $f^0 = \tau-\sigma$, $g^0=\tau+\sigma$.
Then $\vec f'$ and $-\vec g'$ both have unit norm, and as functions of their arguments they trace out closed curves on the unit sphere as $\sigma$ goes around the strings; also, each averages to zero by periodicity.  It follows that these curves will typically intersect (an even number of times).\footnote{A kink implies a discontinuity in $f'$ or $g'$ and so a gap in one of the curves.  Since cosmic strings have a short distance structure with many kinks, there will be many gaps and this may reduce the number of intersections.}
 Each intersection represents an event that occurs once per period, and when it occurs the $\sigma$-parameterization becomes singular.  A representative example for the leading behavior at the singularity (where for simplicity we put the intersection at $\sigma = \tau = 0$) is
\begin{eqnarray}
f^x &=& (\sigma - \tau) - c^2 (\sigma - \tau)^3/2 \ ,\quad 
f^{y} = c (\sigma - \tau)^2 \ ,\nonumber\\
g^x &=& -(\sigma + \tau) + c'^2 (\sigma + \tau)^3/2
\ ,\quad g^{y} = c' (\sigma + \tau)^2 \ ,
\end{eqnarray}
with $c$ and $c'$ constants of order the inverse size of the loop.  It follows (for the generic case that $c \neq c'$) that at $\tau = 0$ the string forms a cusp, $y \propto |x|^{2/3}$.

The instantaneous velocity $(\vec f' - \vec g')/2$ approaches the speed of light at the cusp.  Like the crack of a whip, a great deal of energy is concentrated in the tip, but this whip is perhaps hundreds of light-years long and with a tension not so far below the Planck scale.  It thus emits an intense beam of gravitational waves in the direction of its motion~\cite{Damour:2000wa}.
The Fourier transform of a cusp singularity is much larger at high frequency than for a smooth function, so that even when the suppression for being off-axis is included this comes within reach of LIGO.

This is shown in Fig.~4, reproduced from~\cite{Damour:2001bk}. 
\begin{figure}[t]
\begin{center}
\includegraphics[height=.3\textheight]{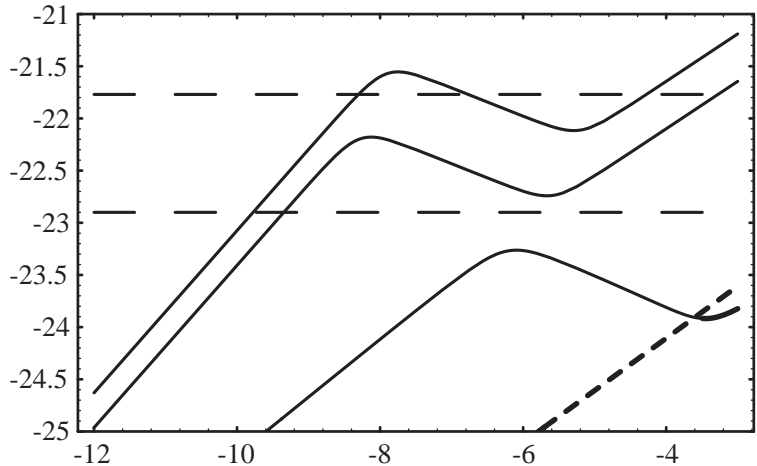}
\end{center}
\caption{Gravitational wave cusp signals, taken from Damour and Vilenkin~\cite{Damour:2001bk}.
The horizontal axis is $\log_{10} \alpha$ where $\alpha = 50G\mu$.  Thus the brane inflation range $10^{-12} \,{\stackrel{<}{{_\sim}}}\, G\mu \,{\stackrel{<}{{_\sim}}}\, 10^{-6}$ becomes $-10.3 < \log_{10} \alpha < -4.3$. The vertical axis is $\log_{10} h$ where $h$ is the gravitational strain in the LIGO frequency band.  The upper and lower dashed horizontals are the sensitivities of LIGO I and Advanced LIGO at one event per year.  The upper two curves are the cusp signal under optimistic and pessimistic network assumptions; the pessimism is that a large number of kinks may suppress the cusps.  The lowest solid curve is the signal from kinks, which form whenever strings reconnect.  The dashed curve is the stochastic signal.}
\end{figure}
 Under optimistic assumptions (but not, I think, too optimistic), even LIGO I is close to discovery sensitivity of one event per year over much of the range of interesting tensions, including the narrow range of the {\mbox{K\hspace{-7pt}KLM\hspace{-9pt}MT}} model.  This is remarkable: cosmic superstrings might be the brightest objects in gravitational wave astronomy, and the first discovered!  LIGO I to date has around $0.1$ design-year of data, but it is supposed to begin a new science run in January 2005 at close to design sensitivity and with a good duty cycle.  Advanced LIGO is sensitive over almost the whole range, and with a higher event rate; it has a target date of 2009.  LISA, a few years later, is even more sensitive.  In magnitude of strain it is comparable to LIGO I, but it is looking at a frequency 10,000 times lower and so the typical cusp strains are 1000 times greater~\cite{Damour:2001bk}.  The cusp events might be seen in a search for unmodeled bursts.  The shape is not as complex as for stellar and black hole inspirals, but modeling the specific frequency dependence will increase the signal-to-noise ratio.  The power-law frequency dependence of the cusp is distinctive.  See also~\cite{Siemens:2003ra} for a discussion of the form of the cusp signal.
 
The dependence of the sensitivity on string tension in Fig.~4 is interesting because it is not monotonic.  This comes about as follows~\cite{Damour:2001bk}.  As the string tension decreases, the coupling to gravity becomes weaker and so does the gravitational wave burst from a given cusp.  However, since gravitational radiation is the only decay channel for string loops they will live longer.\footnote{There is another effect as well, the dependence of the short-distance network structure on the gravitational radiation, but we can overlook this for simplicity.}  Thus as we decrease $G\mu$ there are more but smaller cusps.  With smaller intrinsic events one cannot see as far, but there are more events to see; the competition between these effects depends on the geometry of the universe.  Thus, the three regimes that are evident in Fig.~4 correspond at the smallest tensions to cusps that took place at redshifts less than one, at the intermediate tensions to cusps that are at redshifts greater than one but in the matter-dominated era, and at the largest tensions to cusps that occurred during the radiation-dominated era.  The rise in the event rate with decreasing tension in the middle range comes about because the signals from smaller, later, cusps suffer less from redshifting during the relatively rapid matter-dominated expansion.

The pulsar bounds are also strengthed by taking into account the cusps.
This is analyzed in detail in~\cite{Damour:2004kw}; this quotes a current bound on $G\mu$ of around $10^{-7}$, but the curve is very flat as a function of $G\mu$ so that a small uncertainty in the pulsar analysis or the network properties produces a much larger uncertainty in $G\mu$.  Correspondingly, improved pulsar data and improved understanding of the networks would make it possible to reach much smaller values of $G\mu$.

One may wonder whether the enormous energy stored in cosmic strings can manifest itself in other spectacular effects.  The cusps appear to be the best source potential source of high energy particles and radiation, but~\cite{Blanco-Pillado:1998bv} argues that this is still not observable.  Note that short distance physics affects the string only very near the cusp, and even LIGO frequency range is looking on a much longer scale.

\section{Distinguishing Superstrings}

If gravitational wave cusps or some other signature of cosmic strings are seen, this will be just the beginning of the story.  Detailed observations will be able to determine some of the microscopic properties of the strings.  For example, it is possible to cleanly distinguish weakly coupled fundamental strings from cosmic strings in perturbative GUTs.

\subsection{Reconnection}

The microscopic structure of the string core does not affect the evolution of strings that are light-years in length, except when two strings cross and their cores interact.  We have noted that gauge theory solitons will always reconnect.  For F-strings, reconnection is a quantum process, and takes place with a probability $P$ of order $g_{\rm s}^2$.  The numerical factors are worked out in~\cite{Jackson:2004zg}.  To be precise, $P$ is a function of the relative angle and velocity in the collision, but it is simplest to the value averaged over collision parameters.  

An important issue is the motion of the string in the compact dimensions.   For many supersymmetric compactifications, strings can wander over the whole compact space.  Thus they can miss each other, leading to a substantial suppression of $P$~\cite{Jones:2003da,Dvali:2003zj}.  However, we have noted that in realistic compactifications strings will always be confined by a potential in the compact dimensions.  Even if the scale of the potential is low, the fluctuations of the strings are only logarithmic in the ratio of scales --- this is characteristic of one-dimensional objects~\cite{Copeland:2003bj}.  Thus there is no suppression by powers of the size of the compact dimensions, but the logarithm can be numerically important --- it tends to offset powers of $\pi$ that appear in the numerator.  The value of $g_{\rm s}$, and the scale of the confining potential, are not known, but in a variety of models~\cite{Jackson:2004zg} finds $10^{-3} \,{\stackrel{<}{{_\sim}}}\, P \,{\stackrel{<}{{_\sim}}}\, 1$.  For D-D collisions the situation is more complicated, and in the same models one finds $10^{-1} \,{\stackrel{<}{{_\sim}}}\, P \,{\stackrel{<}{{_\sim}}}\, 1$.  For F-D collisions, $P$ can vary from 0 to 1.

To determine the observational effect of changing $P$ one must feed the given value into the network evolution.  A simple argument suggests that the signatures scale as $1/P$: the amount of string in the network must be increased by this factor in order for an increased number of collisions per unit length of string to offset the reduced $P$ in each collision~(\cite{Damour:2004kw} and references therein).  This is a bit oversimplified, because there are issues connected with the sub-horizon scale structure in the string network~\cite{Bennett:1989yp, Austin:1993rg} that can work in either direction, but is supported by simulations as well~(\cite{Sakellariadou:2004wq} and reference therein).\footnote{Refs.~\cite{Martins:2004vs,Avgoustidis:2004zt} discuss other network evolution issues.  We should note that the second of these is concerned with higher-dimensional excitations which, according to our discussion, should be massive and so decoupled except at early times.}
  If we take this $1/P$ as a model, we see that for the smaller values of $P$ discussed above there can be a substantial increase in the signal even above the encouraging values found in the last section, so that LIGO might soon begin to see {\it many} cusps.  Of course, the existing bounds become stronger.
   
 To first approximation there are two relevant parameters, $\mu$ and $P$.  Each individual cusp event has only a single parameter to measure, its overall strength $h$: because it is a power law  there is no characteristic frequency scale.  (There is a high frequency cutoff, determined by the alignment of the cusp with the detector~\cite{Damour:2001bk}, but this gives no information about the cusp itself.)  After $O(10)$ cusps are seen one can begin to plot a spectrum, $dN \sim A h^{-B} dh$, and from the two parameters $A$ and $B$ constrain $\mu$ and $P$.
There are degeneracies --- $B$ depends primarily on the epoch in which the cusp took place --- but with a more detailed spectrum, and ultimately with data from kink events and pulsars, this degeneracy will be resolved.  Thus $\mu$ and $P$ will be overdetermined, and nonstandard network behavior (such as we are about to discuss) will be detectable.

If $P$ is only slightly less than one, say $0.5$, then it will require precision simulation of the networks and good statistics on the signatures to distinguish this from 1.0.  It should be noted that even with given values of $\mu$ and $P$ there are still substantial uncertainties in the current understanding of the behavior of string networks.  This has recently been discussed in~\cite{Damour:2004kw}, which concludes that the sensitivities given in Fig.\ 3 are only weakly dependent on the unknowns.

The discussion of cosmic strings from grand unified theories has focussed on perturbative unification, as suggested by the successful prediction of the Weinberg angle.  In this case the only strings are the classical solitons, and $P = 1$ is a fairly robust signature.  Thus these can be differentiated from fundamental strings unless we are very unlucky and $g_{\rm s}$ is very close to 1, or we have a very unusual field theory.\footnote{One example: a soliton string with a massless internal scalar mode (besides its normal collective coordinates) would be like a string moving in a higher dimensional space and so two such strings could `miss' each other in field space.  However, such modes will always get a mass from symmetry breaking, and I expect that this effect will be significant only in rather contrived models.}

Strongly coupled confining gauge theories can have electric flux tube strings.  For these the reconnection probability is of order $1/N_{\rm color}^2$, so finding a small  value for $P$ would not rule out a field-theoretic origin entirely, just a perturbative one.  We should note that perturbative string theory gives a prediction for the functional dependence on the collision parameters, $P(v,\theta)$~\cite{Polchinski:1988cn, Jackson:2004zg}.  It would seem very difficult to determine this from observations.  If it were possible to map out the string network in detail, through lensing, then perhaps this function might be studied.

\subsection{Networks}

The second potentially distinguishing feature of the superstring networks is the existence of both F- and D-strings~\cite{Dvali:2003zj, Copeland:2003bj, Jackson:2004zg}, and moreover bound states of $p$ F-strings and $q$ D-strings with the distinctive tension formula (\ref{eq:tensions}).
In this case, when strings of different types collide, rather than reconnecting they form more complicated networks with trilinear vertices.  It is then possible that the network does not scale, but gets into a frozen phase where it just stretches with the expansion of the universe~\cite{Kibble:1976sj,Vilenkin:1984rt}.  If so, its density would come to dominate at the tensions that we are considering.  The F-D networks have not yet been simulated, but simulations of comparable networks suggest that they scale, possibly with an enhanced density of strings~\cite{Vachaspati:1986cc,Spergel:1996ai,McGraw:1997nx}.
From the discussion above, it follows that one will not directly read off the spectrum~(\ref{eq:tensions}) from the observations, but there should eventually be enough information to distinguish F-D string networks from other types.

Warped models with more complicated throats, as well as models with large compact dimensions, can give rise to a richer spectrum of strings~\cite{Sarangi:2002yt, Jones:2003da}.  Still, the $(p,q)$ spectrum~(\ref{eq:tensions}) is worth particular attention: it is an attractive possibility that inflation takes place in a warped throat, and this is the spectrum that arises in the most generic throat.  Thus, while the landscape of string theory may be vast, this particular local geometric feature may be common to a large swath of it.

Networks with multiple types of string can also arise in field theories, though I do not know any {\it perturbative} field theory that gives the particular spectrum~(\ref{eq:tensions}).  However, because of duality there will be gauge theory strings that are very hard to differentiate from the F- and D-strings that we are discussing.  In particular, in the {\mbox{K\hspace{-7pt}KLM\hspace{-9pt}MT}} model the strings exist in a Klebanov-Strassler throat~\cite{Klebanov:2000hb}, which has a dual description as a cascading gauge theory, in which there are $(p,q)$ strings.\footnote{For discussions of relations between field theory strings and F- and D-strings see~\cite{Becker:1995sp,Edelstein:1995ba,Dvali:2003zh,Halyo:2003uu,Binetruy:2004hh,Gubser:2004qj,Achucarro:2004ry,Gubser:2004tf,Lawrence:2004sm}; see~\cite{Rocher:2004uv,Rocher:2004my}
for some related recent bounds.}
  
Indeed, the existence of dualities between string theories and field theories raises the issue, what really is string theory?  This is beyond our current scope, but I note that in the present case there is a quantitative question.  The Klebanov-Strassler throat has a parameter $g_{\rm s}M$; when this is large the string description is the valid one, and when it is small the gauge description is the valid one.  If we can get enough information about the string network,  perhaps combined with information from the CMB, then there might well be enough information to test the hypothesis that we are seeing a Klebanov-Strassler theory, and to measure $g_{\rm s}M$.

\section{Conclusions}

As we have seen, each of the four conditions that we discussed at the beginning is independently model-dependent.  However, quite a number of things have worked out surprisingly well: the production of strings in brane inflation, the possible stabilization of the strings as a side effect of other properties of the models (in particular, of the stabilization of the vacuum itself), the possibility to see strings over many interesting orders of magnitude of tension, and the existence of properties that distinguish different kinds of string so that after the strings are discovered we can do a lot of science with them.  In any case, searching for cosmic strings is a tiny marginal cost on top of experiments that will already be done, and it is great that string theorists will have a stake in these experiments over the coming decade and more.

\subsection*{acknowledgments}
I would like to thank E. Copeland, M. Jackson, N. Jones, and R. Myers for collaborations, and  N. Arkani-Hamed, L. Bildsten, G. Dvali, A. Filippenko, T. Kibble, A. Linde, A. Lo, A. Lommen, 
J. Preskill, S. Trivedi, H. Tye, T. Vachaspati, and A. Vilenkin for discussions and communications.  This work was supported by National Science Foundation
grants PHY99-07949 and PHY00-98395.

\end{document}